\documentclass[11pt,a4paper]{article}
\usepackage[authoryear]{natbib}
\usepackage{times}
\usepackage{graphics}
\usepackage{ifpdf}
\usepackage{graphicx}
\usepackage{amssymb}
\usepackage{amsmath,latexsym,xspace}
\usepackage{array}
\usepackage{setspace} 
\usepackage{comment}
\usepackage{authblk}
\def\Fig#1{Figure~\ref{#1}}

\usepackage[scale={0.73,0.86},centering,includeheadfoot]{geometry}

\def\Prob{\text{Prob}}
\def\E{\text{E}}

\def\Var{\text{Var}}

\def\erfc{\text{erfc}}
\def\mm#1{\ensuremath{\boldsymbol{#1}}}

\def\eref#1{(\ref{#1})}

\setkeys{Gin}{width=0.45\textwidth}
\newcommand{\PCP}{PC prior}

\begin{document}
\begin{titlepage}
\title{Penalised complexity priors for stationary autoregressive processes}
\author[1]{Sigrunn Holbek S\o rbye}
\author[2]{H\aa vard Rue}

\affil[1]{Department of Mathematical Sciences, UiT The Arctic University of Norway, 9037 Troms{\o}, Norway. e-mail: sigrunn.sorbye@uit.no}
\affil[2]{Department of Mathematical Sciences, Norwegian University of Science and Technology, 7491 Trondheim, Norway. e-mail: hrue@math.ntnu.no}
\date{\today}
\end{titlepage}
\maketitle
\begin{abstract}
The autoregressive process of order $p$ (AR$(p)$) is a central model in time series analysis. A Bayesian approach requires the user
to define a prior distribution for the coefficients of the AR$(p)$
model. Although it is easy to write down some prior, it is not at all
obvious how to understand and interpret the prior, to ensure that it
behaves according to the users prior knowledge. In this paper, we approach this problem using the recently developed
ideas of penalised complexity (PC) priors. These priors have important
properties like robustness and invariance to reparameterisations, as well as a clear interpretation. 
A PC prior is  
computed based on specific principles, where model component complexity
is penalised in terms of deviation from simple base model
formulations.  
In the AR$(1)$ case, we discuss two natural base model choices,
corresponding to either independence in time or no change in time. The latter case is illustrated in a survival model with possible
time-dependent frailty. For higher-order processes, we propose a
sequential approach, where the base model for AR$(p)$ is the
corresponding AR$(p-1)$ model expressed using the partial
autocorrelations. The properties of the new prior are compared with the reference prior in a simulation study. 
\end{abstract}

{\bf Keywords:} AR$(p)$, \texttt{R-INLA}, prior selection, robustness.
 
\section{Introduction}
Autoregressive (AR) processes are widely applied to model time-varying stochastic processes, for example within finance, biostatistics and natural sciences \citep{brockwell:02, chatfield:03,prado:10}. Applications also include Bayesian model formulations, often combined with  Markov chain Monte Carlo computations to perform posterior and predictive inference \citep{albert:93, chib:93, barnett:96}.  Particularly, AR processes are useful to model underlying latent dependency structure and they make up important building blocks in complex hierarchical models,  for example analysing spatial data  \citep{lesage:97, sahu:07, sahu:12}. 
 
In fitting an AR($p$) process using a  Bayesian approach, it is necessary to select priors for all model parameters. A simple choice is to  assign uniform priors to the regression coefficients \citep{zellner:71, dejong:91}, but this is not optimal neither for the first-order nor higher-order processes \citep{berger:94}. A more reasonable approach is given by  \cite{liseo:13}, who provide a general framework to compute both Jeffreys and reference priors using the well-known partial autocorrelation function (PACF) parameterisation \citep{barndorff:73}. Stationarity of the AR($p$) process is equivalent to choosing the partial autocorrelations within a $p$-dimensional unit hypercube. In general, Jeffreys priors are invariant to reparameterisations, while reference priors are not. \cite{liseo:13} recommend reference priors, at least when the order of the AR process is smaller or equal to 4. For higher-order processes, calculation of the reference prior is numerically cumbersome and requires extensions of their suggested numerical approximation. 

This paper derives and investigates penalised complexity (PC) priors  \citep{simpson:15} for the  partial autocorrelations of stationary AR processes of any finite order. In general, a PC prior is  computed based on specific underlying principles, in which a model component is seen as a flexible parameterisation of a simple base model structure. The main  idea is to assign a prior to a measure of divergence from the flexible version of the component to its base model and the  PC prior for the relevant parameter is derived by transformation.  In the AR(1) case, this implies that the PC prior for the first-lag coefficient $\phi$  can be derived using white noise ($\phi=0$) as a base model. Alternatively, we can view the limiting random walk case as a base model ($\phi=1$), representing no change in time. Which of these base models that represent a natural choice depends on the relevant application. 

In the higher-order AR($p$) case, we introduce a sequential  approach to construct a PC prior for the $p$th partial autocorrelation, using the corresponding AR$(p-1)$ process as a base model. The resulting joint prior for the partial autocorrelations is  consistent under marginalisation, and each of the marginals can be adjusted according to a user-defined scaling criterion. The scaling is important and prescribes the degree of informativeness of the prior. Here, we suggest to incorporate a scaling  criterion using the variance of the one-step ahead forecast error, also allowing for different rates of shrinkage for each of the partial autocorrelations. The resulting priors have good robustness properties and are also seen to have comparable frequentistic properties with reference priors. 

The plan of this paper is as follows. PC priors and their properties are reviewed in Section~\ref{sec:pc}. We derive PC priors for the coefficient of an AR(1) process in Section~\ref{sec:ar1}, using the two mentioned base models. PC priors are designed to prevent overfitting and this property is demonstrated for a real data example in Section~\ref{sec:coxph}, where an AR(1) process is used to model time-dependent frailty in a Cox proportional hazard model. Contrary to previous results \citep{book114,art551}, the given data on chronic granulomatous disease do not seem to support the additional introduction of a time-varying frailty. Extension of the PC priors to higher-order AR processes is given in Section~\ref{sec:higher}, including incorporation of interpretable scaling parameters to adjust the rate of shrinkage.  Section~\ref{sec:sim} contains simulation results, comparing the performance of the PC and reference priors, while concluding remarks are given in Section~\ref{sec:discussion}. 

\section{Penalised complexity priors and their properties}\label{sec:pc}
The framework of PC priors \citep{simpson:15} represents a systematic and unified approach to compute priors for parameters of model components with an inherit nested structure.  A simple version of the model component is referred to as a base model, typically characterised by a fixed value of the relevant parameter, while the flexible version is seen as a function of the random parameter. The PC prior is computed to penalise deviation from the flexible model to the fixed base model. This section gives a brief review on PC priors and their properties in the context of AR$(p)$ processes.

\subsection{A brief review on the principles underlying PC priors}\label{sec:pc-review}
The informativeness of PC priors is specified in terms of four main principles, stated in \cite{simpson:15}. These principles are useful both to compute priors in a unified way and to understand their properties. The principles, summarised below, express support to Occam's razor, penalisation of model complexity using the Kullback-Leibler divergence, a constant rate penalisation and user-defined scaling, see \cite{simpson:15} for a thorough description of PC priors and their applications.

\begin{enumerate}
\item Let $f_1=\pi(\mm{x}\mid \xi)$ denote the density of a model component $\mm{x}$ in which we aim to find a prior for the parameter $\xi$. A simpler structure of this model component is characterised by the density $f_0=\pi(\mm{x}\mid\xi=\xi_0)$, where $\xi_0$ is a fixed value. In accordance with the principle of parsimony expressed by Occam's razor, the prior for $\xi$ should be designed to give proper shrinkage to $\xi_0$ and decay with decreasing complexity of $f_1$. 

\item In order to characterise the  complexity of $f_1$ compared with $f_0$, we calculate a measure of complexity between these two densities. 
PC priors are derived using the  Kullback-Leibler divergence \citep{kullback:51}, 
\begin{equation*}
\mbox{KLD}(f_1 \parallel f_0)=\int f_1(x)\log\left(\frac{f_1(x)}{f_0(x)}\right)dx, \label{eq:kl-general}
\end{equation*}
which measures the information lost when the flexible model $f_1$ is approximated with the  simpler model $f_0$. For zero-mean multi-normal densities, calculation of the Kullback-Leibler divergence simplifies to performing simple matrix computations  on the covariance matrices as 
\begin{eqnarray*}
\mbox{KLD}(f_1\parallel f_0)  &= & \frac{1}{2}\left(\mbox{tr}(\mm{\Sigma}_0^{-1} \mm{\Sigma}_1)-n-\ln\left(\frac{|\mm{\Sigma}_1|}{|\mm{\Sigma}_0|}\right)\right)\label{eq:kl}
\end{eqnarray*}
where 
$f_i\sim N(0,\mm{\Sigma}_i)$, $i=0,1$, while $n$ is the dimension. 
To facilitate interpretation, the Kullback-Leibler divergence is transformed to a unidirectional distance measure
\begin{equation}
d(\xi)=d(f_1\parallel f_0)= \sqrt{2\mbox{KLD}(f_1 \parallel f_0)}.\label{eq:distance}
\end{equation}
This is not a distance metric in the ordinary sense, but a quantity which is interpretable as a measure of distance from the flexible model $f_1$ to the base model $f_0$.
\item  In choosing a prior for the distance measure $d(\xi)$, it is natural to assume that the mode should be located at the base model while the density decays as the distance from the base model increases. The PC prior is derived based on a principle of constant rate penalisation, 
\begin{equation}
\frac{\pi(d(\xi)+\delta)}{\pi(d(\xi))}=r^{\delta},\quad d(\xi),\delta\geq 0, \label{eq:rate}
\end{equation}
where $r\in (0,1)$. This implies that the relative change in the prior for $d(\xi)$ is independent of the actual distance. 
This is a reasonable choice as it is complicated to properly characterise different decay rates for different distances. 
The resulting prior is exponentially distributed, $\pi(d(\xi))=\lambda\exp(-\lambda d(\xi))$, $\lambda>0$, where $r=\exp(-\lambda)$ and the corresponding PC prior for $\xi$  follows by a standard change of variable transformation. 

\item The rate $\lambda$ characterises the shrinkage properties of the prior and it is important that this parameter can be chosen (implicitly) in an intuitive and interpretable way, for example by a user-defined probability statement for the parameter of interest. \cite{simpson:15} suggest to determine $\lambda$ by incorporating a probability statement of tail events, e.g. 
\begin{equation}
P(Q(\xi)>U)=\alpha,\label{eq:upper}
\end{equation}
where $U$ represents an assumed  upper limit for an interpretable transformation $Q(\xi)$,  while $\alpha$ is a small probability. However, other scaling suggestions might be just as reasonable, depending on the specific application.  
\end{enumerate}

\subsection{Important properties of PC priors in the context of AR processes}
The given four principles provide a strategy to calculate priors for model parameters in a systematic way, rather than turning to ad-hoc prior choices still often made in Bayesian literature. Also, the principles can be helpful to interpret the assumed prior information and how this influences posterior results.

A first important property of PC priors is invariance to reparameterisations. This follows automatically as the prior is derived based on a measure of divergence between models, which does not depend on the specific model parameterisation. We consider the invariance property to be particularly useful in the case of autoregressive processes, as these are typically parameterised either in terms of the regression coefficients, or by using the partial autocorrelations. The great benefit of using the partial autocorrelations is that these give an unconstrained set of parameter values, ensuring a  positive definite correlation matrix. In contrast, the valid parameter space for the regression coefficients is rather complicated, especially for higher-order processes ($p>3$). 
 
Second, the PC priors are designed to shrink towards well-defined base models. In the setting of autoregressive processes, this implies that the priors will  prevent overfitting, for example in terms of selecting an unnecessarily high order of the process. In addition, the base model can be chosen to reflect different simple structures of a model component, depending on the given application. For an AR(1) process, it is relevant to assume either no dependency, or no change in time, as simple base model formulations. For higher-order processes, we could also choose no correlation as a base model but this might cause too much shrinkage in many applications. As an alternative, we introduce a new sequential approach which defines a sequence of base models, reflecting the additional complexity in increasing the order of the fitted AR process. 

Third, PC priors are computationally simple and are already implemented within the  \texttt{R-INLA} framework \citep{rueal:09, martins:13}, for different latent Gaussian model components. The priors are designed to have a clear interpretation as the informativeness of the priors is adjusted by user-defined scaling. Here, we will take advantage of this to allow for different rates of shrinkage for priors assigned to partial autocorrelations of different lags.   In contrast, objective priors simply aim to incorporate as little information to the inference as possible. 

\section{PC priors for AR(1) using two different base models}\label{sec:ar1}
A  first-order autoregressive process can be defined by  
\begin{equation*}
x_t=\phi x_{t-1}+w_t,\quad w_t\sim N(0,\kappa^{-1}), \quad t=2,\ldots , n,\label{eq:ar1}
\end{equation*}
where $x_1$ is assumed to be normally distributed with mean 0 and marginal precision $\tau=\kappa(1-\phi^2)$.  This  process represents an important special case of general autoregressive processes, in which the dependency structure is governed by the coefficient $\phi$. Using the framework of penalised complexity priors, $\phi$ is viewed as a flexibility parameter reflecting deviation from simple fixed base model formulations. In this section, we derive PC priors for $\phi$ both using no autocorrelation ($\phi=0$) and no change in time ($\phi=1$) as base models, and we suggest how these priors can be scaled. A real-data application using the latter base model is included in Section~\ref{sec:coxph}. 

Note that we also use a penalised complexity prior for the precision parameter $\tau$.  Following \cite{simpson:15},  this prior is derived using infinite precision as a base model, which gives the type-2 Gumbel distribution 
\begin{equation}
\pi(\tau) =\frac{\lambda}{2}\tau^{-3/2}\exp(-\lambda \tau^{-1/2}),\quad \lambda>0. \quad \label{eq:prec}
\end{equation}
The rate $\lambda$ is inferred using the probability statement $P(1/\sqrt{\tau}>U)=\alpha$, where $\alpha$ is a small probability. 
The prior is scaled by specifying  an upper limit $U$ for the marginal standard deviation $1/\sqrt{\tau}$, in which  the corresponding rate is  $\lambda=-\log(\alpha)/U$. To make an intuitive choice for $U$, one can consider the marginal standard deviation after the precision $\tau$ is integrated out. For example, if $\alpha=0.01$ this standard deviation is $0.31U$ \citep{simpson:15}. 

\subsection{Base model: No dependency in time}
In general, the correlation matrix of the first-order autoregressive process is  $\mm{\Sigma}_1 = \left(\phi^{|i-j|}\right)$. Choosing no autocorrelation ($\phi=0$) as a base model, the resulting process is white noise with correlation matrix equal to the identity matrix,  $\mm{\Sigma}_0=\mm{I}$. By simple matrix calculations, the  distance function (\ref{eq:distance}) is seen to equal  $d(\phi)=\sqrt{(1- n)\log(1-\phi^2)}.$  Using the principle of constant rate penalisation (\ref{eq:rate}), an exponential prior  is assigned to $d(\phi)$ with rate $\lambda = \theta/\sqrt{n-1}$. The resulting prior is invariant to $n$ and by the ordinary transformation of variable formula, the PC prior for the one-lag autocorrelation is 
\begin{equation}
    \pi(\phi) = \frac{\theta}{2} \exp\left(-\theta \sqrt{-\ln(1-\phi^{2})}\right) 
    \; \frac{|\phi|}{(1-\phi^2) \sqrt{-\ln(1-\phi^2)}}, \qquad |\phi|<1, \theta>0.\label{eq:base0}
\end{equation}

The rate  parameter $\theta$ is important as it influences how fast the prior shrinks towards the white noise base model.
To infer $\theta$, we need a sensible criterion which facilitates the interpretation of this parameter. \cite{simpson:15} suggest to use a probability statement for an interpretable transformation of the parameter of interest, for example in terms of  tail events as defined by (\ref{eq:upper}). When the base model is $\phi=0$, a reasonable alternative is to define such a tail event as large absolute
correlations being less likely, i.e. 
\begin{equation*}
\Prob(|\phi| > U ) = \alpha. \label{eq:upper-base0}
\end{equation*}
This implies that  $\theta = -\ln(\alpha)/\sqrt{-\ln(1-U^{2})}$. The interpretation of this criterion is intuitive in the first-order case, but we find it difficult to use in practice for higher-order processes. An alternative scaling idea is presented in Section~\ref{sec: scale-prederr}, where we consider the variance of the one-step forecast error as the order of the AR process is increased. We recommend the latter approach, as this is more intuitively implemented for general AR$(p)$ processes. 

\subsection{Base model: No change in time}\label{seq:base1}
An alternative base model for the AR(1) process is to assume that the process does not change in time ($\phi=1$). This represents a limiting random walk case, being a non-stationary and singular process. Consequently, a limiting argument is needed to derive the PC prior for $\phi$.  

Let $\mm{\Sigma}_1 =
\left(\phi^{|i-j|}\right)$ and $\mm{\Sigma}_0 =
\left(\phi_0^{|i-j|}\right)$ where $\phi_0$ is close to $1$ and $\phi
< \phi_0$.  In this case, the Kullback-Leibler divergence is 
$$
\mbox{KLD}(f_1(\phi)\parallel f_0) = \frac{1}{2}\left(\frac{1}{1-\phi_0^2}(n-2(n-1)\phi_0\phi+(n-2)\phi_0^2)-n-(n-1)\ln\left(\frac{1-\phi^2}{1-\phi_0^2}\right)\right).
$$
Considering the limiting value as $\phi_0\rightarrow 1$, the distance 
$$d(\phi) = \lim_{\phi_0\rightarrow 1} \sqrt{2\mbox{KLD}(f_1(\phi)\parallel f_0)} =\lim_{\phi_0\rightarrow 1}\sqrt{\frac{2(n-1)(1-\phi)}{1-\phi_0^2}}=c\sqrt{1-\phi},\quad |\phi|<1,$$ for a 
constant $c$ that does not depend on $\phi$.
Since $0\le d(\phi) \le c\sqrt{2}$, we assign a truncated exponential
distribution to $d(\phi)$ with rate $\theta=\lambda/c$ and the resulting PC prior for $\phi$ is 
\begin{equation}
    \pi(\phi) = \frac{\theta \exp\left(-\theta\sqrt{1-\phi}\right)}{
        \left(1-\exp\left(-\sqrt{2}\theta\right)\right) \;
        2\sqrt{1-\phi}},\quad |\phi|<1. \label{eq:base1}
\end{equation}

Again, we need to suggest an intuitive criterion to scale the prior in terms of $\theta$. This case requires separate consideration, as it cannot be seen as a special case of the approach in Section~\ref{sec:higher}. One option is to make use of  (\ref{eq:upper}),  and determine $(U, \alpha)$ in terms of the probability statement 
$\Prob(\phi > U) = \alpha$. The solution to this equation is given implicitly by
\begin{equation*}
    \frac{1-\exp\left(-\theta\sqrt{1-U}\right)}{
        1-\exp\left(-\sqrt{2}\theta\right)}
    = \alpha, \label{eq:scaling-rho1}
\end{equation*}
provided that $\alpha$ is larger than the lower limit
$\sqrt{(1-U)/2}$.

\subsection{The PC priors versus the reference prior}
\begin{figure}[h]
\begin{center}
\rotatebox{270}{\includegraphics[width=8cm]{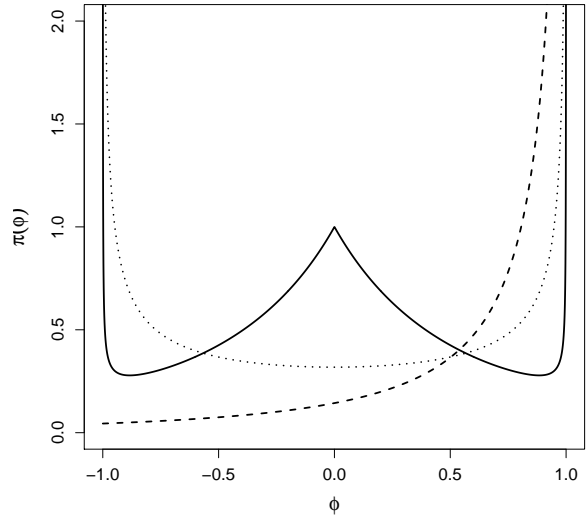}}
\caption{The PC priors for the coefficient $\phi$ of AR(1), using $\phi=0$ (solid thick line) and $\phi=1$ (dashed line) as base models. The rate parameters $\theta$ in (\ref{eq:base0}) and (\ref{eq:base1}) are set equal to 2 in both cases. For comparison we also include the reference prior for $\phi$ (dotted line).}
\label{fig:priors}
\end{center}
\end{figure}

The two alternative PC priors for the first-lag coefficient of an AR(1) process are illustrated in \Fig{fig:priors},  using rate parameter $\theta=2$ in (\ref{eq:base0}) and (\ref{eq:base1}), respectively. For comparison, we also illustrate the reference prior defined by $\pi(\phi) = \frac{1}{\pi}(1-\phi^2)^{-1/2}$, $|\phi|<1$ \citep{barndorff:73,berger:94,liseo:13}. 

In general, reference priors are designed to give objective Bayesian inference in the sense of being least informative in a certain information-theoretic sense \citep{berger:09}. This implies that the data are given a maximum effect on the posterior estimates. In general, the reference prior is calculated to maximise a measure of divergence from the posterior to the prior. In the given AR(1) case, the reference prior for $\phi$ is calculated to maximise an asymptotic version of the expected Kullback-Leibler divergence, in practice performed using an asymptotic version of the Fisher information matrix \citep{barndorff:73, liseo:13}. The resulting reference prior is seen to be similar to the Jeffreys prior which is defined (up to a constant) by the square root of the determinant of the Fisher information matrix \citep{liseo:13}. Using a small rate parameter, the PC prior with base $\phi=0$ will be quite similar to the reference/Jeffreys prior but for increasing rate parameters, the effect of shrinkage to 0 is increased. Note that a PC prior using $\phi=-1$ as the base model can be derived similarly as using the $\phi=1$ base model.

\section{Application: Modeling time-varying frailty with AR(1)}
\label{sec:coxph}
To demonstrate the use of the PC prior for the lag-one autocorrelation, we consider an example of a Cox proportional
hazard model with time-varying frailty. The Cox proportional hazard model is a popular type of survival model
that can be fitted to recurrent event data. It assumes that the
time-varying hazard for the $i$th subject can be expressed as
$h(t; i) = h_0(t) \exp(\eta_i)$,
where the combined risk variable $\eta_i$ in most cases depends on
subject-specific covariates $\mm{z}_i$ and contributions from random
effects/frailty. The function $h_0(t)$ is the baseline hazard,
see~\citet{book114} for further details and applications of the
model. In the given example, our main focus is on the inclusion of a
subject-specific and possibly time-dependent frailty term in $\eta_i$.

\subsection{Dependent Gaussian random effects}
\label{sec:dependentGRE}

A full Bayesian analysis of the Cox proportional hazard model requires a model for the baseline
hazard. A natural choice is to consider the log baseline hazard as a
piecewise constant function on small time intervals, and impose
smoothness to penalise deviations from a constant, see for
example~\citet[Sec~8.1.1]{book65} and \citet[Sec.~3.3.1]{book80}.  Let
$[0, T]$ be the time interval of interest, and divide that interval
into $n$ equidistant (for simplicity) intervals $0 < t_1 < t_2 <
\cdots < t_{n-1} < T$. Let $h_j$, $j=1, \ldots, n$ denote the log
baseline hazard in the $j$th interval. The first order random walk
(RW1) model imposes smoothing among neighbour $h_i$'s,
\begin{displaymath}
    \pi(\mm{h} \mid \tau) \propto
    (\tau\tau^{*})^{(n-1)/2} \exp \left( -\frac{\tau\tau^{*}}{2}
      \sum_{j=2}^{n} (h_j - h_{j-1})^{2} \right).
\end{displaymath}
This is a first-order intrinsic Gaussian Markov random field with a covariance matrix on the form $\tau^{-1} R$, where the correlation matrix $R$ is singular and of rank $n-1$. The parameter $\tau^{*}$ is a positive scaling constant which is added such that the generalised variance (the geometric mean of the diagonal elements of $R^{-1}$),  is 1. This is needed to make the model invariant to the size of $n$ and to unify the interpretation of $\tau$, which then represents the precision of the (marginal) deviation from the null space of $R$, see  \cite{sorbye:13} and \cite{simpson:15} for further details. To separate the baseline hazard from the intercept, we
impose the constraint $\sum_i h_i = 0$. The base model is a constant
(in time) baseline hazard, which corresponds to infinite smoothing,
$\tau = \infty$. The resulting penalised complexity prior for $\tau$ is given by (\ref{eq:prec}). 

An interesting extension to the commonly used subject specific frailty
model is to allow the frailty term to depend on time~\citep{art551},
leading to a time-dependent combined risk variable $\eta_i(t)$.
Anticipating a positive correlation in time, it is natural to model
this time dependent risk using a continuous-time Ornstein-Uhlenbeck
process or its discrete time version given by AR(1). The stationary
AR(1) model for subject $i$'s specific frailty is given by
\begin{displaymath}
    v_{it} \mid \{v_{is}, s < t\} \;\sim \;{\mathcal N}(\phi
    v_{i,t-1}, 1/(\tau_v (1-\phi^{2}))),
\end{displaymath}
parameterised so that $\tau_v$ is the marginal precision and $\phi$
is the lag-one correlation. For this model component, the natural base model (keeping the marginal
precision constant) is a time-constant frailty, in which we use the  PC prior for $\phi$ in (\ref{eq:base1}).
For a fixed correlation $\phi$, the base model for the precision $\tau_v$ is 
the constant zero which gives the type-2 Gumbel prior in~(\ref{eq:prec}).

\subsection{Analysis of chronic granulomatous disease data}
\label{sec:cgd}
We end this section by analysing data on chronic granulomatous disease
(CGD)~\citep{book114} available in \texttt{R} as the \texttt{cgd}
dataset in the \texttt{survival} package.  This data set consists of
$128$ patients from $13$ hospitals with CGD. These patients
participated in a double-blinded placebo controlled randomised trial,
in which a treatment using gamma interferon ($\gamma$-IFN) was used to
avoid or reduce the number of infections suffered by the patients.
The recorded number of CGD infections for each patient ranged from
zero to a maximum of seven, and the survival times are given as the
times between recurrent infections on each patient. We
follow~\citet{art551} and introduce a deterministic time dependent
covariate for each patient, given as the time since the first
infection (if any). Additionally, we include the covariates treatment
(placebo or $\gamma$-IFN), inherit (pattern of inheritance), age (in
years), height (in cm), weight (in kg), propylac (use of prophylactic
antibiotics at study entry), sex, region (US or Europe), and steroids
(presence of corticosteroids) \citep{art550,art551}. The covariates
age, height and weight were scaled before the analysis.  

The computations were performed using the \texttt{R-INLA} package, by
rewriting the model into a larger Poisson regression,
see~\citet{book65} for a more general discussion and~\citet{art494}
for \texttt{R-INLA} specific details. The prior specifications are as
follows. We used a constant prior for the intercept and independent
zero mean Gaussian prior with low precision, i.e. $0.001$, for all the
fixed effects.  For the log baseline hazard with $n=25$ segments, we
used the type-2 Gumbel prior with parameters $(U=0.15/0.31,
\alpha=0.01)$ giving a marginal standard deviation for the log
baseline hazard of about $0.15$. This seems adequate as we do not
expect the log baseline hazard to be highly variable.  The
time-dependent frailty was assigned a type-2 Gumbel prior for the
precision with parameters $(U=0.3/0.31, \alpha=0.01)$ giving a
marginal standard deviation of about $0.3$, hence we allow for
moderate subject specific variation. For the derived
prior~(\ref{eq:base1}) for $\phi$, we used the parameters $(U = 1/2,
\alpha = 0.75)$, which puts most of the prior mass for high values of $\phi$ as  $P(\phi > 1/2) = 0.75$. This corresponds to using a rate parameter $\theta\approx 1.55$ in (\ref{eq:base1}). 

\begin{figure}[h]
    \centering \mbox{
        \rotatebox{270}{\includegraphics[width=6cm]{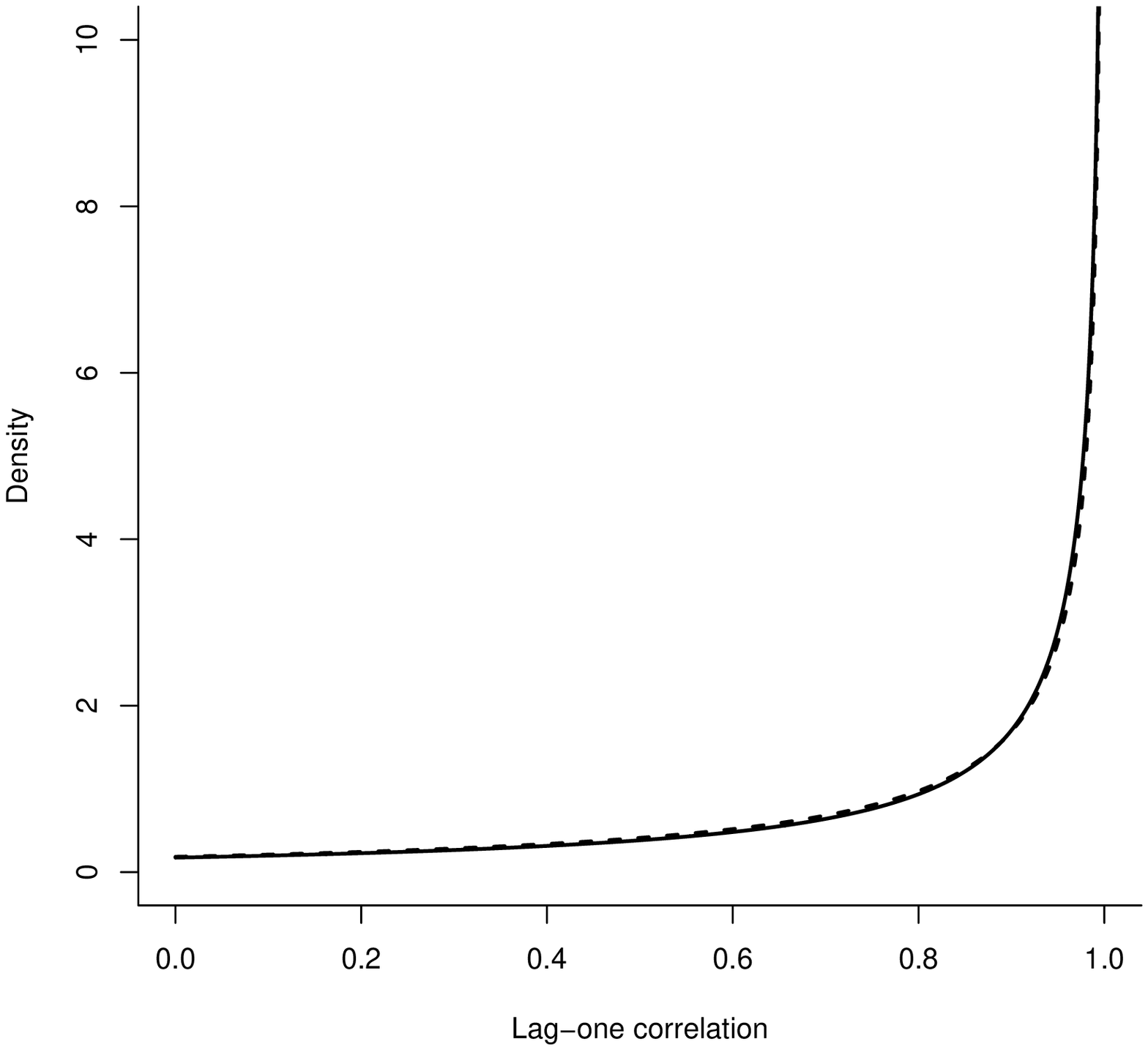}}
        \makebox[5mm]{}
        \rotatebox{270}{\includegraphics[width=6cm]{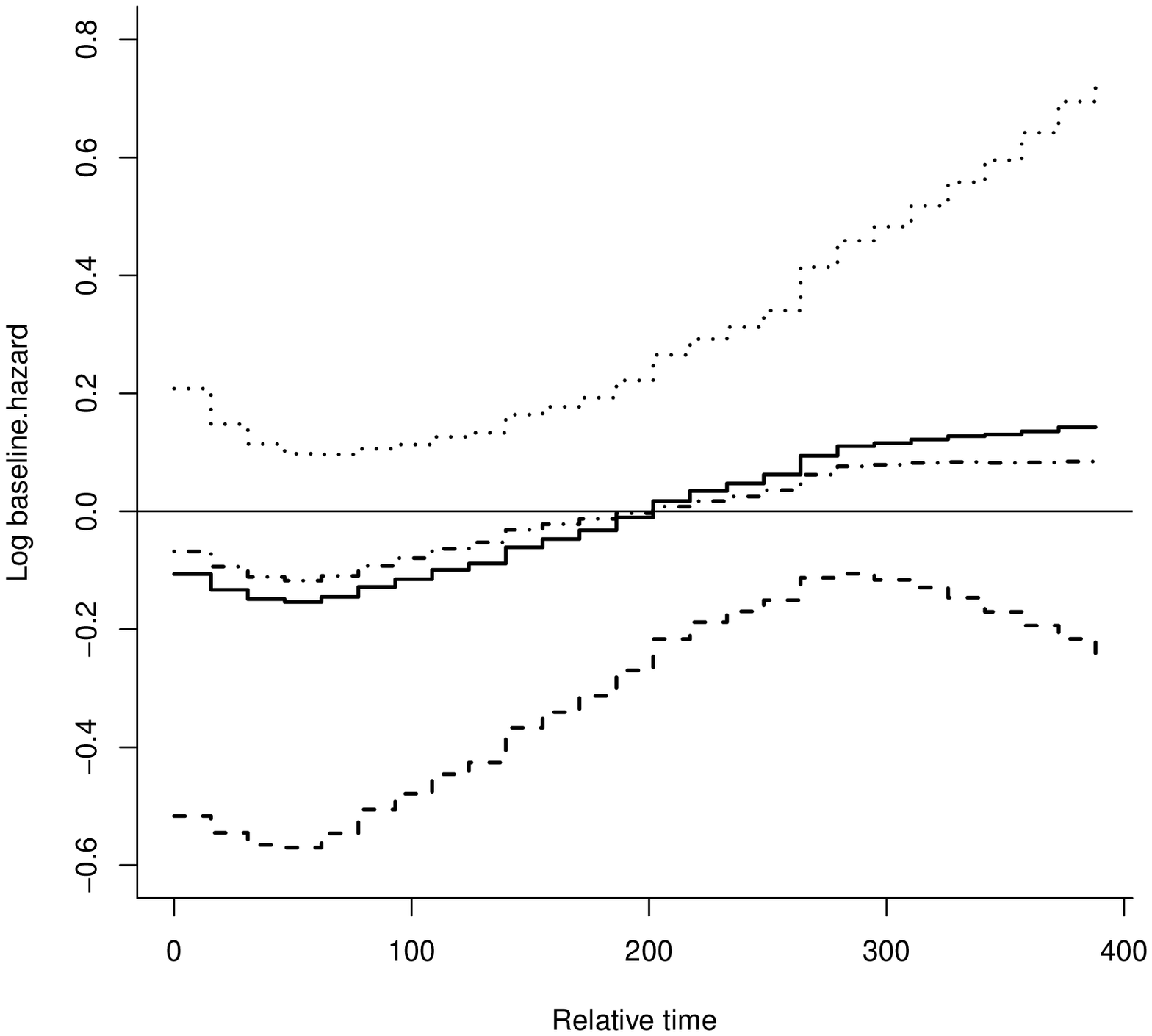}}}
    \mbox{\makebox[14.5cm]{}} \mbox{
        \makebox[6cm]{(a)}
        \makebox[5mm]{}
        \makebox[6cm]{(b)}}
    \caption{Panel~(a) displays the posterior density (solid) and
        prior density (dashed) for the lag-one autocorrelation $\phi$ in
        the AR(1) model for the time-dependent frailty.  Panel~(b)
        displays the log baseline hazard, mean (solid), median
        (dashed-dotted), lower ($0.025$, dashed) and upper ($0.975$,
        dotted) quantiles.}
    \label{fig:coxph1}
\end{figure}

\Fig{fig:coxph1}~(a) shows the prior (dashed) and posterior
 (solid) densities for the autocorrelation coefficient of the AR(1) model for the
frailty. The data hardly alters the prior at all, showing that there
is not much information in the data available for this parameter, and
we cannot conclude anything about the time-varying frailty. This is
contrary to the findings in~\citet{art550} and
\cite{art551}. \Fig{fig:coxph1}~(b) displays the log baseline hazard,
showing an increasing trend (additional to the deterministic time
dependent covariate), but the wide point-wise credible bands give no
clear evidence for a time-dependent baseline hazard.  With the new
prior we are more confident that we do not overfit the data using the
more flexible model for the log baseline hazard, as we do control the
amount of deviation and its shrinkage towards it. The given
conclusions are robust to changes in the parameter choices
$(U,\alpha)$ for the different model components.

\section{Deriving PC priors for higher-order AR processes}\label{sec:higher} 
Define an autoregressive process of order $p$   by  
\begin{equation}\label{corr:eq1}%
    x_t = \phi_1 x_{t-1} + \cdots + \phi_p x_{t-p} + \epsilon_t,\quad  \quad \epsilon_t \stackrel{\text{iid}}{\sim} {\mathcal N}(0, \kappa^{-1}),
\end{equation}
where  $\mm{x}=(x_1,\ldots , x_n)$ is an $n$-dimensional vector, $t=p,\ldots, n$, and $\kappa$ is the precision of the innovations. The corresponding $p\times p$ correlation matrix $\mm{\Sigma}_p$ is
Toeplitz~\citep{book56} with elements that can be expressed as  $\Sigma_{ij} = \sigma_{|i-j|}$,  where $\sigma_0=1$. Although~\eref{corr:eq1} is a natural parameterisation for known parameter values 
$\mm{\phi}_p=(\phi_1,\ldots ,\phi_p)$, it is an awkward parameterisation when these are unknown, as the positive definiteness   requirement of the correlation matrix makes the space of valid
$\mm{\phi}_p$ complicated for $p > 3$. This implies that it is necessary to impose a number of non-linear constraints on  these coefficients to define a stationary process.

A good alternative is to make use of the 
invariance property of the \PCP\ and define the prior for $\mm{\phi}_p$
implicitly.  The basic idea, which is commonly used when estimating
AR$(p)$ parameters, is to assign the prior to the partial autocorrelations $\mm{\psi}_p=(\psi_1,\ldots , \psi_p) \in
[-1,1]^{q}$, where $q = p-1$. This gives a useful unconstrained set of
parameters for this problem.  Furthermore, there is a smooth bijective mapping
between the partial autocorrelations and the autocorrelations  in $\mm{\Sigma}_p$, given by the Levinson-Durbin recursions
\citep{monahan1984note,book2}. 

\subsection{A sequential approach to construct PC priors}
In deriving PC priors for the partial autocorrelations of an AR$(p)$ process, we suggest to use a sequential approach, augmenting the partial autocorrelations one by one. Define $\psi_0=0$  and assume that $\mm{\psi}_p=(\mm{\psi}_{p-1},\psi_p)$ for  $p=1,2,\ldots .$
We calculate the Kullback-Leibler divergence, conditional on the terms already included in the model,
\begin{eqnarray*}
\mbox{KLD}(f_1(\mm{\psi}_p) \parallel f_0(\mm{\psi}_{p-1}))  =  \frac{1}{2}\left(\mbox{tr}(\mm{\Sigma}_{p-1}^{-1} \mm{\Sigma}_{p})-n-\ln\left(\frac{|\mm{\Sigma}_{p}|}{|\mm{\Sigma}_{p-1}|}\right)\right), 
\label{eq:kl2}
\end{eqnarray*}
where $\mm{\Sigma}_0=\mm{I}$ and $f_1$ and $f_0$ represent the densities of the  AR$(p)$ and AR$(p-1)$ processes, respectively. Notice that by augmenting the partial autocorrelations $\mm{\psi_{p-1}}$ with one (or several) terms, the correlation structure between the first $p-1$ elements of the corresponding AR$(p)$ process remains unchanged. As the inverse correlation matrix of  the  AR$(p-1)$ process is a  band matrix of order $2p-1$, we immediately notice that  
$$\mm{\Sigma}_p^{-1} \mm{\Sigma}_{p+r}=\mm{I},\quad r=1,2\ldots,$$
and $\mbox{tr}(\mm{\Sigma}_{p-1}^{-1} \mm{\Sigma}_{p})=n$.  Also, 
\begin{eqnarray*}
 \ln\left(\frac{|\mm{\Sigma}_{p}|}{|\mm{\Sigma}_{p-1}|}\right) &=& \ln\left(\frac{\prod_{i=1}^{p}(1-\psi_i^2)^{n-i}}{\prod_{i=1}^{p-1}(1-\psi_i^2)^{n-i}}\right)
 = (n-p)\ln(1-\psi_{p}^2).
\end{eqnarray*}

The resulting measure of distance from the AR$(p)$ model to its base AR$(p-1)$, is only a function of the $p$th order partial autocorrelation, i.e.,
$$d(\psi_{p}) =\sqrt{2\mbox{KLD}(f_1(\mm{\psi}_p) \parallel f_0(\mm{\psi}_{p-1}))}  =  \sqrt{-(n-p)\ln(1-\psi_{p}^2)}.$$
Applying the principle of constant rate penalisation \eqref{eq:rate}, an exponential density is assigned to $d(\psi_{p})$ with rate $\lambda_{p}=\theta_p/\sqrt{n-p}$. The resulting prior for the $p$th partial autocorrelation is  
\begin{equation}
 \pi(\psi_{p}) = \frac{\theta_p}{2} \exp\left(-\theta_p \sqrt{-\ln(1-\psi_{p}^{2})}\right) 
    \; \frac{|\psi_{p}|}{(1-\psi_{p}^2) \sqrt{-\ln(1-\psi_{p}^2)}}, \qquad |\psi_{p}|<1,\label{eq:prior-pacf}
\end{equation}
where the parameter $\theta_p>0$ influences how fast the prior shrinks towards the base model.

The given formulation allows us to derive interpretable conditional priors for  each of the partial autocorrelations $\psi_p$, given the previous parameters $\mm{\psi}_{p-1}$. As the resulting priors are conditionally independent, the partial autocorrelations are seen to be consistent under marginalisation (as discussed in \cite{west:91} in the context of kernel density estimation). Also, the  marginal for an AR$(q)$ process is not influenced by higher-order partial autocorrelations when these are 0, i.e. for $q\leq p$:
\begin{eqnarray*}
\pi(\mm{\psi}_{q})&=&\int \pi(\mm{\psi}_p)d\mm{\psi}_{-q}= \pi(\mm{\psi}_q \mid \psi_{q+1}=0,\ldots , \psi_{p}=0). 
\end{eqnarray*}

\subsection{Controlling shrinkage properties} \label{sec: scale-prederr}
The given sequential approach implies that the prior for partial autocorrelations of different lags have the same functional form, but potentially different rate parameters. The next step is to determine a reasonable criterion to choose the rate $\theta_p$  in (\ref{eq:prior-pacf}).  Our suggestion is motivated  by the conditional variance of the one-step ahead forecast error for an AR($p$) with fixed $p$, 
$$\Var\left(( x_{t+1}-\hat x_{t+1})\mid \mm{x}_{s\leq t},\tau\right)=\tau^{-1} (1-\psi_1^2)(1-\psi_2^2)\cdots (1-\psi^2_p),$$
and the observation that often $1-\psi^2_k$ is an non-decreasing function with $k$. We assume that
\begin{equation*}
\E(1-\psi_k^2)=1-(1-a)b^{k-1}, \quad a,b \in [0,1],\quad k=1,\ldots , p,\label{eq:a-gamma}
\end{equation*}
so the one-step ahead prediction, a priori, is non-decreasing with $k$.  This reduces the prior specification into two parameters $a$ and $b$,  which have to be specified by the user. The parameter $a$ represents the initial expectation $\E(1-\psi^2_1)=a$. The choice $b=1$ induces the same shrinkage for all $\psi_k$ while $b <1$ gives increasing shrinkage for increasing $k$. For given values of  $a$ and $b$, the corresponding value for the rate parameter  in (\ref{eq:prior-pacf}) is  found by solving  
\begin{equation}
\E(1-\psi_k^2)=\frac{\theta_k\sqrt{\pi}}{2}\exp\left(\frac{\theta_k^2}{4}+\log\left(\erfc\left(\frac{\theta_k}{2}\right)\right)\right) = 1-(1-a)b^{k-1}
\label{eq:solve-ab}
\end{equation}
for each $k=1,\ldots , p$, where $\erfc(z)$ denotes the complementary error function
$$\erfc(z)=\frac{2}{\sqrt{\pi}}\int_{z}^{\infty}e^{-t^2}dt.$$

\section{Simulation results} \label{sec:sim}
To illustrate the properties of PC priors for the partial autocorrelations of autoregressive processes, we conduct a simulation study in which an AR(3) process is fitted to six different test cases. Except for the two first cases, the test examples are similar to the ones used in \cite{liseo:13}.  In each case, we fit an AR(3) model to  generated time series of length $n=50$.  
\begin{table}[h]
\begin{center}
\begin{tabular}{|l|rrr|rrr|}\hline
 & \multicolumn{3}{c|}{Root mean squared error} & \multicolumn{3}{c|}{Coverage ($95\%$)} \\
Test cases &$\widehat{\mbox{rmse}}_1$& $\widehat{\mbox{rmse}}_2$&$\widehat{\mbox{rmse}}_3$& $\hat \zeta_1$ & $\hat \zeta_2$& $\hat \zeta_3$   \\\hline
PC prior ($a=b=0.5$) & & & & & & \\
1. $\mm{\psi}=(0,0,0)$ & 0.133 & 0.123 & 0.111 & 0.928 & 0.919 & 0.956 \\
2. $\mm{\psi}=(0.7,0,0)$ &  0.106 & 0.118 & 0.103 & 0.912 & 0.961 & 0.968\\  
3. $\mm{\psi}=(0.2,0.3,0)$ &  0.174 & 0.150 & 0.106 & 0.888 & 0.882 & 0.968\\  
4. $\mm{\psi}=(-0.2,-0.6,0)$ &  0.070 & 0.123 & 0.108 & 0.953 & 0.921 & 0.959\\  
5. $\mm{\psi}=(0.5,-0.3,0)$ & 0.093 & 0.136 & 0.107 & 0.938 & 0.918 & 0.965\\  
6. $\mm{\psi}=(0.5,-0.3,-0.1)$ & 0.092 & 0.146 & 0.118 & 0.937 & 0.892 & 0.950\\  \hline\hline
Reference prior & & & & & &\\
1.  $\mm{\psi}=(0,0,0)$ &0.146 & 0.151 & 0.135 & 0.911 & 0.901 & 0.931 \\
2.  $\mm{\psi}=(0.7,0,0)$ &  0.101 & 0.143 & 0.126 &0.911& 0.932 & 0.944\\ 
3.  $\mm{\psi}=(0.2,0.3,0)$  & 0.185&  0.143 &0.130 &  0.879 & 0.920 & 0.929\\
4.  $\mm{\psi}=(-0.2,-0.6,0)$ & 0.070 & 0.111 & 0.133 & 0.949 & 0.933 &0.934\\ 
5.  $\mm{\psi}=(0.5,-0.3,0)$ & 0.092 & 0.133 & 0.130 & 0.939 & 0.923 & 0.938\\ 
6. $\mm{\psi}=(0.5,-0.3,-0.1)$ & 0.088 & 0.143 & 0.133 & 0.938 & 0.916 & 0.928\\\hline
\end{tabular}
\caption{The root mean squared error and the frequentistic coverage of $95\%$ highest posterior density intervals for each of the estimated partial autocorrelations of AR(3) processes, using PC priors with $a=b=0.5$ and the reference prior, respectively.  The given results are averaged over 1000 simulations, and the time series length in each simulation is $n=50$.}
\label{tab:sim}
\end{center}
\end{table}

The results using $m=1000$ simulations, are displayed in Table~\ref{tab:sim}, where the average root-mean squared error  is
$$\widehat{\mbox{rmse}}_i=\sqrt{\frac{1}{m}\sum_{j=1}^m (\hat\psi_i-\psi_i)^2},\quad i=1,2,3.$$
We also report frequentistic coverage, $\hat \zeta_i$, $i=1,2,3,$ of  the estimated $95\%$ highest posterior density intervals. In all test examples, the PC prior was implemented with scaling $a=b=0.5$. By solving  (\ref{eq:solve-ab}), this corresponds to using rate parameters $(\theta_1,\theta_2,\theta_3)\approx (0.87, 1.94, 3.33)$ in estimating the three partial autocorrelations.

As expected, the results illustrate that the use of PC priors avoid overfitting. In the first test case of simulating white noise, the PC prior is seen to give both smaller root mean squared error  and better frequentistic coverage, compared with using the reference prior. We also notice that using PC priors gives better results in estimating $\hat \psi_3$ for all the test cases. For the other parameters, the PC and reference priors are seen to have quite comparable performance. This implies that the PC prior seems like a promising alternative to reference priors in estimating the partial autocorrelations of AR($p$) processes. The main advantage of PC priors is that these are easy to compute, also for higher-order processes, and more flexible than the reference prior, allowing for individual scaling. In comparing the two priors, we also considered the forecast error and coverage of $95\%$ highest posterior density intervals for one-step ahead predictions. The results were very similar using the PC and reference priors and we do not report these here.

The given results are not surprising. Especially, the approach of scaling the PC priors is designed to reflect decreasing partial autocorrelations as the order of the process is increased.  If we have reasons to believe that the partial autocorrelations do not decrease with higher order, we suggest to scale the priors for the partial autocorrelations of all lags similarly, using $b=1$. We have chosen to report results only using $a=b=0.5$ but have also tried several other combinations of the scaling parameters $a$ and $b$. The main impression is that the PC priors are quite robust to different choices of $a$ and $b$. Also, it is easy to understand how changes in these parameter will induce changes in the estimates. A larger value of $a$ and/or smaller value of $b$ give more shrinkage to 0. In general, we recommend that  $a$ is chosen to be less or equal to 0.5 as higher values of $a$ might impose too must shrinkage for the first-lag partial autocorrelation. Also, values of $b$ less than 0.5 might impose too much shrinkage for the partial autocorrelations of higher lags.

\section{Discussion}\label{sec:discussion}
An important aspect of statistical model fitting is to select models that are flexible enough to capture true underlying structure but do not overfit.  Among competing models we would prefer the more parsimonious one, for example in terms of having fewer assumptions, fewer  model components or a simpler structure of model components. \cite{hawkins:04} describes overfitting in terms of violating the principle of parsimony given by Occam's razor, the models and procedures used should contain all that is necessary for the modeling but nothing more. The given PC priors obey this principle, ensuring shrinkage to specific base models chosen to reflect the given application.

The PC priors represent a weakly informative alternative to existing prior choices for autoregressive processes, allowing for user-defined scaling to adjust the informativeness of the priors. The PC priors are computationally simple and are easily implemented for any finite order $p$ of the autoregressive process. The priors  are available within the $\texttt{R-INLA}$ framework, in which AR processes can be used as building blocks within the  general class of  latent Gaussian models \citep{rueal:09}. This class of models have many applications, among others including analysis of temporal and spatial data. A natural extension in time series applications is to derive PC priors also for autoregressive (integrated) moving average processes. Other useful model extensions would include vector autoregressive models \citep{sims:80}, frequently used to analyse multivariate time series, for example within the fields of econometrics. 

In this paper, we have only considered the stationary case. Previous controversy \citep{phillips:91} in assigning a prior to the lag-one autocorrelation of an AR(1) process relates to whether the stationarity condition $|\phi|<1$ is included, or not.  \cite{phillips:91} argued that objective ignorance priors, like the Jeffreys prior, should be used for AR(1) processes when no stationarity assumptions are made, while uniform priors would give  inference biased towards stationarity. One of the problem seen with Jeffreys prior is that it puts most of its probability mass on regions of the parameter space giving a  non-stationary process \citep{liseo:13}. The reference prior was originally only defined for stationary process but has been extended in a symmetric way for $|\phi|>1$ \citep{berger:94}, in which it is seen to have a more reasonable shape than Jeffreys prior \citep{robert:07}. A relevant future project is to study the use of PC priors also for non-stationary AR processes.

\bibliographystyle{apa}
\bibliography{shs}    
\end{document}